\documentclass[cits]{PoS}
\newcommand{\be}{\begin{equation}}
\newcommand{\ee}{\end{equation}}
\newcommand{\bqn}{\begin{eqnarray}}
\newcommand{\eqn}{\end{eqnarray}}
\title{Higgs mechanism in five-dimensional gauge theories\footnote{BUW/07-09}}
\ShortTitle{Higgs mechanism in five-dimensional gauge theories}
\author{\speaker{Magdalena Luz} and Francesco Knechtli\\
        Bergische Universit\"at Wuppertal\\
        E-mail: \email{luz@physik.uni-wuppertal.de}, 
                \email{knechtli@physik.uni-wuppertal.de}}
\author{Nikos Irges\\
        University of Crete\\
        E-mail: \email{irges@physics.uoc.gr}}

\abstract{Lattice simulations of five-dimensional gauge theories on an 
orbifold revealed that there is spontaneous symmetry breaking. Some of the 
extra-dimensional components of the gauge field play the role of a Higgs 
field and some of the four-dimensional components become massive gauge
bosons. 
The effect is confirmed by computing the Coleman-Weinberg potential with a 
cutoff. We compare the results of this computation with the lattice data.}

\FullConference{The XXV International Symposium on Lattice Field Theory\\
		 July 30-4 August 2007\\
		 Regensburg, Germany}

\begin{document}
\section{Introduction}
Gauge theories in more than 4 dimensions have gained a lot of attention as a possible
generalization of the gauge~-~Higgs sector of the Standard Model. 
Many different models have been discussed in the literature, one of the common 
features being that
the extra dimensions are taken to be
compact with a compactification scale $1/R$.
The expectation is that this leads to a four-dimensional effective theory at a
scale $E \ll 1/R $ whose zero modes correspond to the
Standard Model particles.
In this effective theory, some components of the gauge field in the extra 
dimension take on the role of the Higgs particle
and the Higgs potential is 
generated dynamically through quantum corrections 
\cite{Coleman:1973jx}.
From the four-dimensional point of view these fields act as scalars and 
can potentially acquire a vacuum 
expectation value. That is, the gauge symmetry breaks spontaneousely via 
the Hosotani mechanism \cite{Hosotani:1983xw} and the 
gauge particles become massive just as in the Standard Model.
Whether this occurs in a given model has to be determined by 
examining the Higgs potential in each case.

We study a system with
$SU(N), N=2,3$ on an orbifold $\mathbb{R}^4 \times S^1/\mathbb{Z}_2$.
The five-dimensional fields are expanded in Fourier or Kaluza-Klein (KK) 
modes along the extra
dimension, 
\bqn
      E(x,x_5) & = &  \frac{1}{\sqrt{2\pi R}}E^{(0)}(x) +
      \frac{1}{\sqrt{\pi R}}\sum_{n=1}^\infty
      E^{(n)}(x)\cos(nx_5/R) \quad \mbox{ for even fields}\\ 
      O(x,x_5) & = &  \frac{1}{\sqrt{\pi
           R}}\sum_{n=1}^\infty O^{(n)}(x)\sin(nx_5/R)       \quad \mbox{for odd fields}.
\eqn
The orbifold boundary conditions are implemented in the following way
\cite{Hebecker:2001jb, Irges:2004gy}:
fields related by a reflection of the fifth coordinate are identified up to a
global group conjugation
\be
\begin{array}{l}
g A_\mu(x, x_5) g^{-1} = A_\mu (x, -x_5)\\
g A_5(x, x_5) g^{-1} = -A_5(x, -x_5)
 \end{array}
\mbox{ where }  g^2 \in \mbox{ center of } SU(N).  
\ee
The fixed points of the reflection at $x_5 = 0, \, \pi R$ define 
four-dimensional
boundaries where the gauge group is broken down to a subgroup 
which depends on the choice of $g$.
The even components of $A_5(x)$ transform in some 
representation of the remnant gauge group generated by the even components of $A_\mu(x)$.
For our examples we have
\be
\begin{array}{lll}
SU(2)\stackrel{\mathbb{Z}_2}{\to} U(1) &\mbox{ with } 
            g = -i \sigma^3 & \mbox{ even fields: } 
            A_5^{1,2} \ (Higgs),\   A_\mu^{3} \ (Z)  \\
SU(3)\stackrel{\mathbb{Z}_2}{\to} SU(2) \times U(1) &\mbox{ with }  
           g = {\rm diag}(1, 1, -1) & \mbox{ even fields: } 
            A_5^{4, 5, 6, 7} (Higgs), 
            A_\mu^{1,2,3,8}  ({W^\pm}, Z,\gamma).
\end{array}
\ee
The $SU(3)$ model is the simplest case which generates the electro-weak
symmetry pattern of the Standard Model. There, the Higgs field transforms in the
fundamental representation of the remnant $SU(2)$.

If the scalar field has a non vanishing vev 
the terms involving $A_5$ in the gauge Lagrangean 
     \be
      {\cal L}  =  -\frac{1}{2g_5^2}{\rm tr} \{ F_{\mu \nu}F_{\mu \nu}\} -
                  \frac{2}{2g_5^2}{\rm tr}\{ F_{\mu 5}F_{\mu 5}\}
                  -\frac{1}{g_5^2\xi}{\rm tr}\{ (\bar{D}_M A_M)\}^2
    \ee
generate a mass term for the gauge fields and the scalars through the operator 
$\bar{D}_5\bar{D}_5$, where $\bar{D}_M$ is defined by 
$\bar{D}_M F = \partial_MF+[\langle A_M\rangle,F]$. 
In the $SU(2)$ model, it has 
the eigenvalues 
\bqn
    A_\mu^{3, (0)} \ (Z):& (m_Z R)^2 = \alpha^2 \label{m_z}\\
    A_5^{1, 2 (0)} \ (Higgs):  & (m_{A_5} R)^2 = \alpha^2, 0 \label{m_h}\\
   \mbox{higher KK modes}: & (m_nR)^2 =\frac{n^2}{R^2}, \frac{(n\pm \alpha)^2}{R^2}.
\label{m_kk}
\eqn 
$\alpha$ is related to the vev of the scalar field by
\be
\label{alpha}
\alpha = g_5 \langle A_5^1\rangle R
\ee
and its numeric value is determined by the minimum of the Higgs potential.
A perturbative calculation to one loop yields
\cite{Kubo:2001zc}
\be
\label{v}
    V = -\frac{9}{64\pi^6R^4}\sum_{m=1}^{\infty}\frac{\cos(2\pi
      m\alpha)}{m^5}.
\ee
The minimum of $V$ is at $\alpha = 0$ and as a consequence the remnant gauge symmetry is unbroken and the gauge
particles are massless.
The same is true in $SU(3)$ which suggests that one has to fall back on a more
complicated model if one still hopes to reproduce the Standard Model.
\section{Lattice simulations and perturbation theory at finite cutoff}
However in order to fully explore the viability of extra-dimensional gauge
theories an analysis beyond 1-loop perturbation theory is needed. 
The reason for this is that removing the cutoff in perturbation theory drives
the extra-dimensional gauge
theory to the trivial point. This can be
seen by the following argument:
the theory is parametrized by two dimensionless quantities 
\be
N_5 = \pi R \Lambda, \quad \beta = \frac{2N}{g_5^2 \Lambda}.
\ee
$N_5$ is the ratio of the cutoff $\Lambda$ to the compactification scale
(here we take the interval length $\pi R$) and
$\beta$ the dimensionless coupling which we use in the lattice simulations.
In a perturbative calculation factors of $g_5^2 \Lambda$ can appear from loop
corrections\footnote{For some quantities like the Higgs 
  potential such factors are absent at 1-loop. It is conjectured that 
  there the perturbative series organizes
  itself in powers of the four-dimensional coupling $g_4 = g_5/\sqrt{2\pi R}$ 
  instead. However, this requires to consider the full renormalization as for
  instance at 2-loop logarithmic corrections appear \cite{vonGersdorff:2005ce}.
  For the Abelian theory compactified on $S^1$ a 2-loop calculation 
  has recently been done in
\cite{Hosotani:2007kn} and confirms the conjecture.}.  
Hence, when the cutoff $\Lambda$ is taken to infinity
the dimensionless coupling $g_5^2 \Lambda$ has to vanish,  
in order
to keep the theory perturbative.  
This is only possible where $g_5 \to 0 $ (and $\beta \to \infty$) and the
interactions vanish. 
On the other hand an extra
dimension of finite size $R$ and infinite cutoff also means that $N_5 \to \infty$.
It is therefore only possible to study the truly interacting theory in a
framework where the cutoff is finite and the coupling not necessarily
perturbative.
The lattice provides such a setup.

\begin{figure}
 \begin{minipage}[t]{.49\linewidth}
 \includegraphics[width=1.05\linewidth]{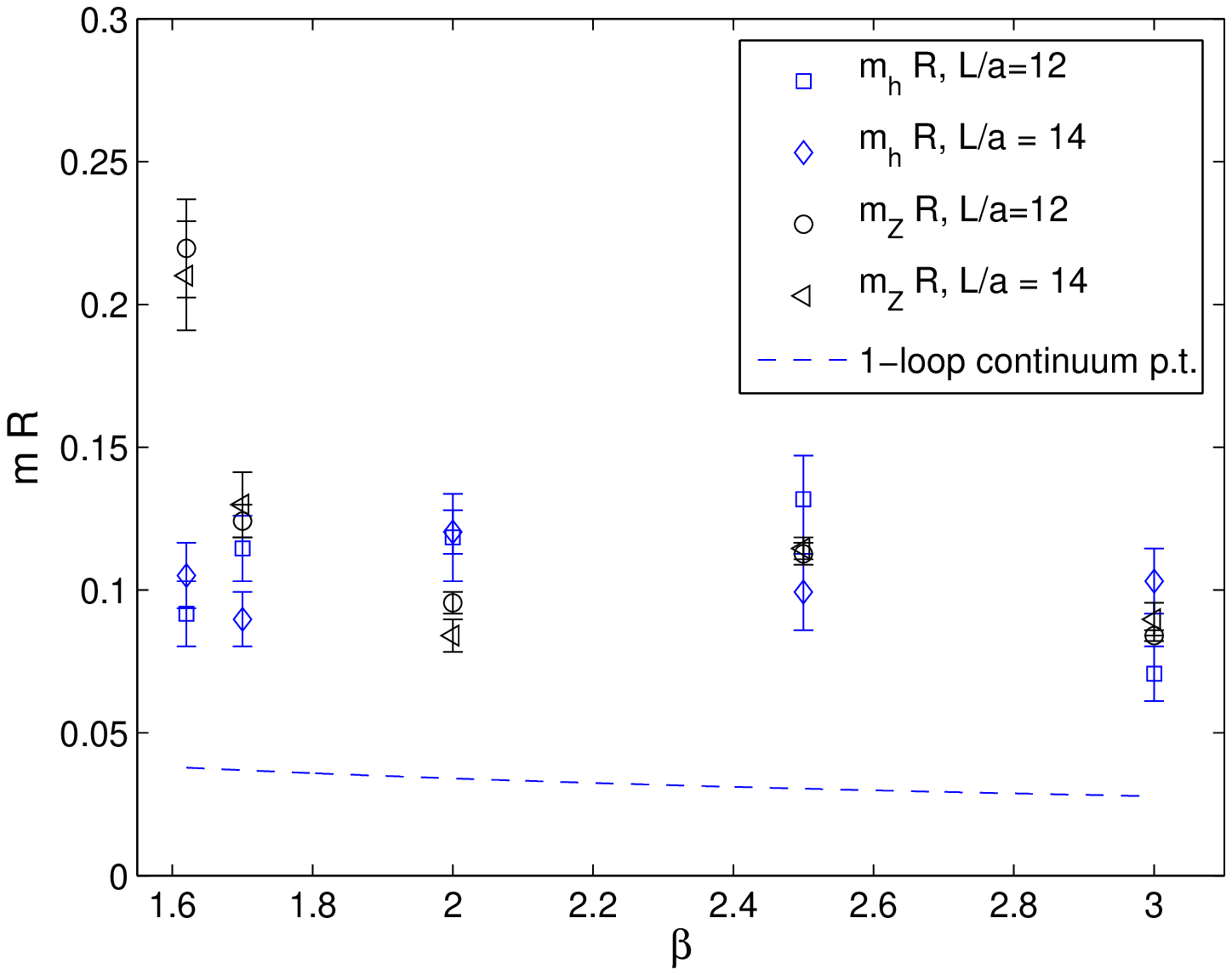}
 \caption{Ground state masses of scalar (square, diamonds) and gauge particles
  (circles, triangles) for $L/a =
  12, 14$ lattices. The dashed line is the 1-loop perturbation theory
  prediction for the Higgs mass.}
\label{masses}
 \end{minipage}
\hfill
\begin{minipage}[t]{.49\linewidth}
\includegraphics[width=1.05\linewidth]{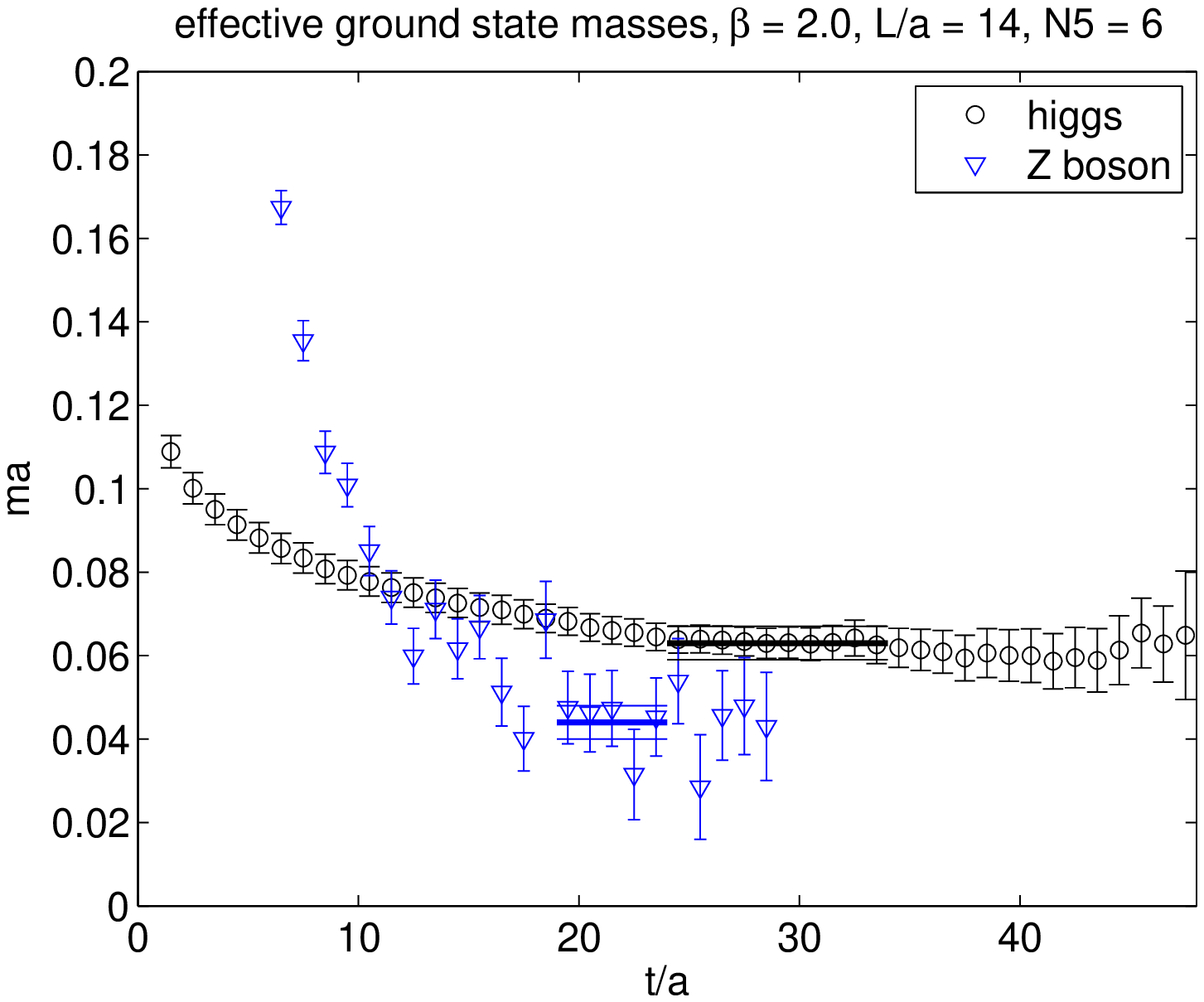}
\caption{Effective masses of the gauge and the scalar particle at $\beta = 2.0
$ from the $L/a = 14 $ lattice. The plateaus where the masses are extracted
are indicated by horizontal bars.}
\label{effmass}
\end{minipage}
 \end{figure}

And indeed,
in contrast to the perturbative results, lattice simulation of the $SU(2)$
model \cite{Irges:2006zf} reveal that the $Z$ boson is massive (cf. Figs.~(\ref{masses},
\ref{effmass})) and for $N_5 = 6$ the Higgs mass is significantly heavier than predicted by
perturbation theory \cite{vonGersdorff:2002as,Kubo:2001zc,Cheng:2002iz}.
The simulations were done on $(T/a) \times (L/a)^3 \times N_5$ lattices where
$a = \Lambda^{-1}$ is the lattice spacing. We use the Wilson plaquette gauge
action.
The system has a first order phase transition 
at $\beta = \beta_c(N_5, L/a)$ which
separates a confined ($\beta < \beta_c$) from a deconfined ($\beta > \beta_c$)
phase. The particle spectra can only be extracted in the latter.
Fig.~(\ref{masses}) shows ground
state masses of the scalar and the gauge boson for different values of the
coupling $\beta$. Finite volume effects are negligible as can be seen from
the figure by
comparing the data from $L/a = 12$ and $L/a = 14$ lattices (both simulations
have $N_5 = 6$ and $T/a = 96$).
In Fig.~(\ref{effmass}) we give an example of the effective masses of the 
two particles at $\beta = 2.0$.

In order to resolve the conflict between the results from perturbation
theory and lattice we redo a perturbative calculation, but leave a finite
cutoff in place.
This can be achieved by describing the lattice action with an effective continuum Lagrangean \`a la
Symanzik. More details on this calculation can be found in \cite{Irges:2007qq}.
The expansion in the lattice spacing is consistently truncated at $O(a^2)$. 
Up to this
order, there are two additional operators which contribute to the mass matrix for the
gauge particles
   \bqn
             c\,{\cal O}^{(6)}   &=& 
     \sum_{M,N} \frac{c}{2}{\rm tr}\{F_{MN}(D_M^2+D_N^2)F_{MN}\}, \qquad c\equiv c^{(6)}(N_5,\beta)\\
           c_0\,{\cal O}^{(5)}   & = & \frac{\pi a{c}_0}{4}
     F_{5\mu}^{\hat a}F_{5\mu}^{\hat a} \left[\delta(x_5) + \delta(x_5-\pi R)\right]
     \,,\qquad c_0\equiv c^{(5)}(N_5,\beta).
   \eqn
${\cal O}^{(6)}$ is a correction from the bulk action 
and ${\cal O}^{(5)}$ is introduced by
the orbifold reflection on the boundary. The coefficients $c$ and $c_0$ are
cutoff dependent through $\beta$ and $N_5$. For the Wilson plaquette action $c
= \frac{1}{12}$ at tree level.
As a consequence, the mass eigenvalues are modified and the shape of the Higgs 
potential changes quite significantly. More concretely, in the $SU(2)$ case
the gauge boson masses change from Eqs.~(\ref{m_z}, \ref{m_kk}) to
 \bqn
\label{m_z_a2}
 A_\mu^{3(0)} \mbox{('Z' boson)}:\;  (m _Z R)^2 &=& \alpha^2  +\frac{c_0
   \alpha^2}{2}\frac{\pi}{N_5} + c\alpha^4 \frac{\pi^2}{N_5^2} \\
 \mbox{higher KK modes: } \; (m_n R)^2 &=& n^2, \qquad n > 0\\
          &=&  (n\pm \alpha)^2 +\frac{c_0 \alpha^2}{2}\frac{\pi}{N_5} + c(n\pm
          \alpha)^4 \frac{\pi^2}{N_5^2},\qquad n \geq 0
\label{kkmass_a2}
 \eqn
where we have truncated the results at $O(a^2)$ and $O(1/n)$. 
\begin{figure}
\includegraphics[width=1.0\textwidth]{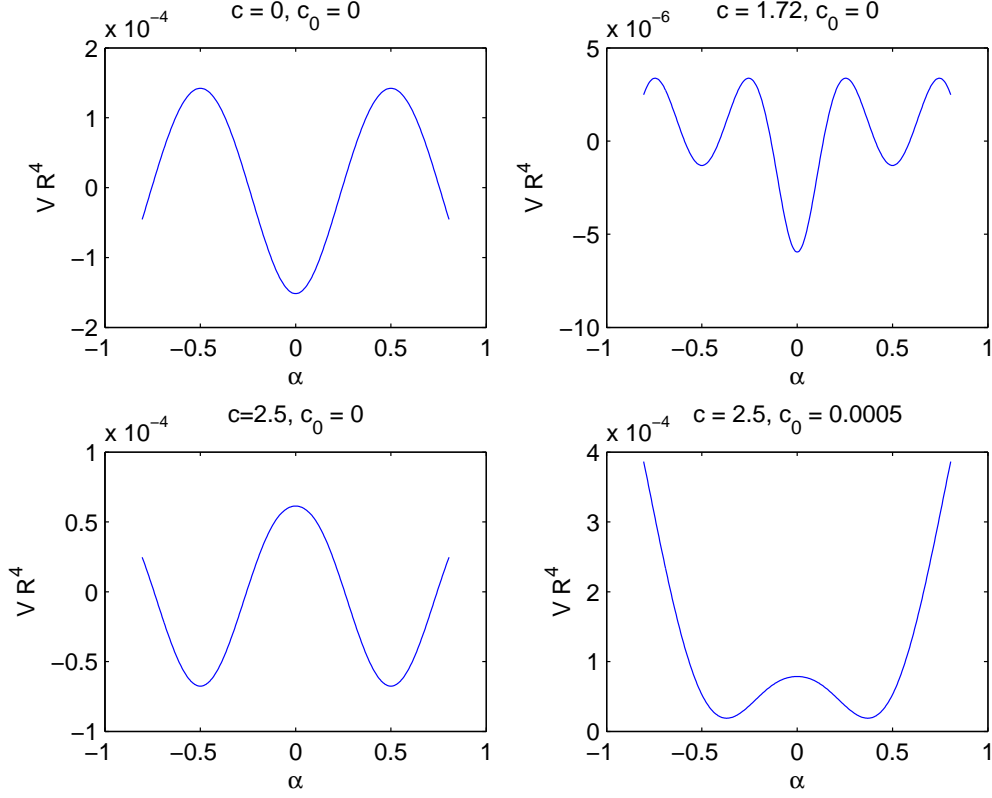}
\caption{1-loop Coleman Weinberg potential: The first plot shows the standard
  perturbative calculation at $\Lambda \to \infty$, the second plot shows
  influence of the bulk cutoff effects for $c = 1.72$. An additional local
  minimum is visiible at  $\alpha = 0.5$, however the global minimum is still
  at $\alpha = 0$. In the third figure at $c = 2.5$ the minimum at $\alpha =
  0 $ has disappeared, the symmetry is spontaneously broken. Finally in the
  last picture the effect of the  boundary coeffiecient $c_0$ appears, the
  minima have shifted away from $\pm 0.5$ to $\pm 0.37$.}
\label{pot}
\end{figure}
The masses of the scalars come from the gauge fixing term and are unchanged
with respect to Eqs.~(\ref{m_h}, \ref{m_kk}). The resulting
Higgs potential depends on the two coefficients $c$ and $c_0$. Some examples
are shown in Fig.~(\ref{pot}). The upper left plot shows the original
potential from Eq.~(\ref{v}) without any cutoff effects ($c = c_0 = 0$). 
If we turn on the bulk effects only, by increasing $c$ but keeping $c_0 = 0$,
a second local minimum appears 
at $\alpha = \pm 0.5$. For large enough $c \geq 1.75$
this minimum turns into a global one, indicating symmetry breaking.
Further increasing $c$ transforms the minimum at $\alpha = 0 $ into a maximum
(cf. the upper right plot and lower left plot in Fig.~(\ref{pot})).
With only the effect of the bulk corrections, it is however not possible 
to shift the miminum of the potential away from either $0$ or $0.5$. 
For this the boundary coefficient $c_0$ is needed as shown in the last plot
of  Fig.~(\ref{pot}). 
The orbifold boundary condition breaks the periodicity of the potential 
and the minimal value of the Higgs potential can be moved continuousely away
from $0$ by varying $c_0$. The finite cutoff immediately also introduces 
a constraint on the value of the vev which should not exeed $1/a$ or
\be
|\alpha| < \sqrt{\frac{N N_5}{\pi^2 \beta}}.
\ee 
 \begin{figure}
 \begin{minipage}[t]{.49\linewidth}
\includegraphics[width=1.0\linewidth]{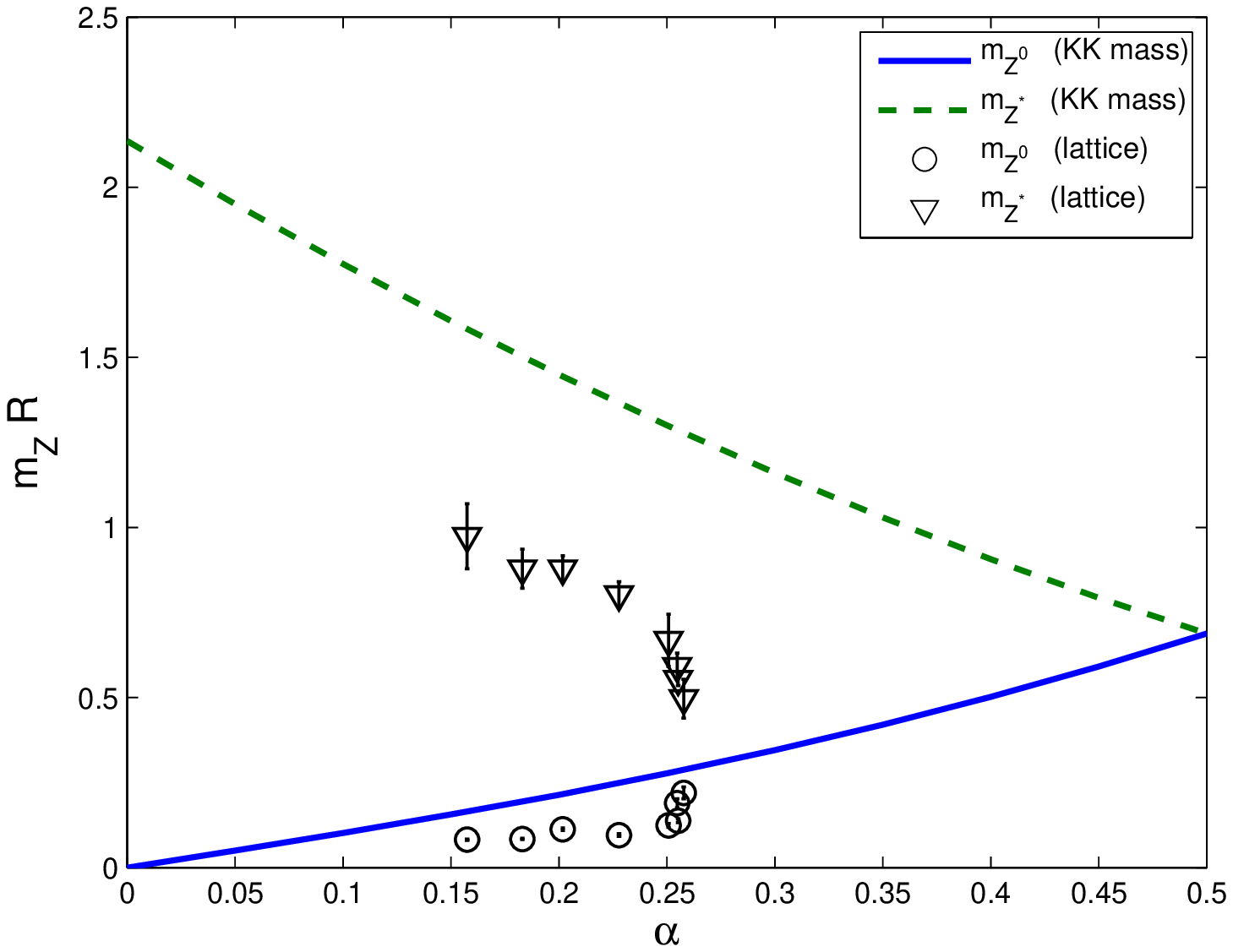}
\caption{Comparison of the gauge boson masses from the lattice with the KK
  masses. The solid line is the ground state of the $Z$ boson, the dashed
line the first excited state. Lattice results (symbols) are at
$L/a=12,\  T/a = 96$.}
\label{mzalpha}
 \end{minipage}
\hfill
\begin{minipage}[t]{.49\linewidth}
\includegraphics[width=1.0\linewidth]{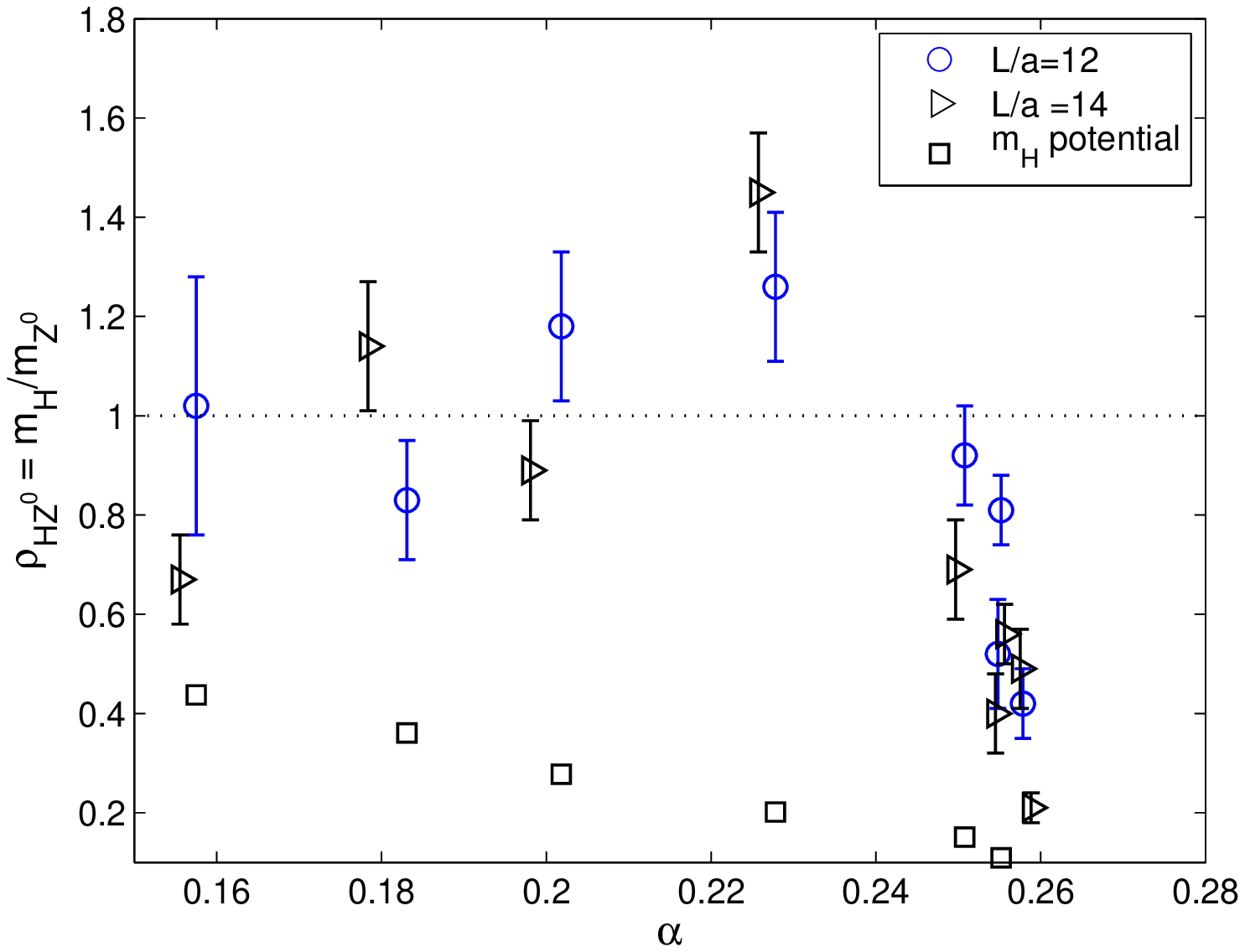}
\caption{Ratio of $m_H$ to $m_Z$. Lattice data from $L/a = 12$ lattices (circles) and $L/a = 14$ lattices (triangles). The squares show the
corresponding ratio from the potential calculation.}
\label{rho}
\end{minipage}
 \end{figure}
In the case of the $SU(2)$ model we can directly compare the cutoff corrected 
potential calculation to our simulation results. Fig. (\ref{mzalpha}) shows the  KK
masses from Eq.~(\ref{m_z_a2}) (ground state, solid line) and 
Eq.~(\ref{kkmass_a2}) for $n = 1$ (first
excited state, dashed line) together with the corresponding lattice masses.
The matching of the lattice to the perturbative setup is done by defining 
\be
\alpha_{lat}(\beta) = \sqrt{\frac{\langle {\rm tr}\{\Phi \Phi^\dag\}\rangle N_5^2 }{2 \pi}}
\ee
where $\langle {\rm tr} \{\Phi \Phi^\dag\}\rangle$ was calculated on the lattice.
The Higgs field $\Phi$ is defined by the commutator of the extra-dimensional
potential $(A_5)_{lat}(n_\mu)$ with the orbifold projection matrix $g$
\be
\phi(n_\mu) \equiv [a (A_5)_{lat}(n_\mu), g] \quad
\mbox{ where } \quad
a (A_5)_{lat}(n_\mu) = \frac{1}{4N_5} (P - P^\dag) 
\ee
and $P$ is the Polyakov line along the extra dimension at the point 
with the four-dimensional integer coordinates $n_\mu$ \cite{Irges:2006hg}. 
We equate $\alpha_{lat}$
with the perturbatively defined $\alpha$ from Eq.~(\ref{alpha}).
Even though we cannot claim quantitative agreement, we do find a similar
qualitative behavior of the perturbative KK masses and their lattice 
counterparts.
In Fig.~(\ref{rho}) we show 
the ratio of the Higgs to the gauge boson mass $\rho_{HZ}$. 
Here, the
matching is done by tuning the coefficients $c$ and $c_0$ in the potential 
such that it takes its minimal value at $\alpha_{min} = \alpha_{lat}$. 
We then compute the Higgs mass from the potential by
\be
     (m_H R)^2 = \frac{N}{N_5\beta}R^4 \left.
\frac{{\rm d}^2 V}{{\rm d}\alpha^2}\right|_{\alpha_{min}}.
\ee
The most striking result in this figure ist that $\rho_{HZ} > 1$ can be
reached on the lattice,
whereas the perturbative results are all way below one, 
\section{Conclusions}
We have calculated the effective Higgs potential in five-dimensional
 pure $SU(N), \ N=2,3$ gauge theory
compactified on an orbifold. In contrast to prior such
results \cite{Kubo:2001zc}, 
we include a finite cutoff explicitly into our calculation. The cutoff effects
are controlled by two coefficients $c$ and $c_0$.  
We find that 
cutoff effects can trigger spontaneous symmetry breaking for 
both $SU(2)$ and $SU(3)$. 
We therefore make contact between the perturbative results and the non
perturbatively defined lattice study where in the case of $SU(2)$ 
massive gauge bosons where found. 

In the case of $SU(3)$ we find that there are combinations of the cutoff
coefficients which lead to the experimentally measured value of the
Weinberg angle of $\cos \theta_w \approx 0.877$, whereas without 
including cutoff effects the value is $\cos \theta_W = 0.5$.
Furthermore it is possible
to obtain $\rho_{HZ^0} > 1$ for small $N_5$. From our point of view these
results are promising and a fully non perturbative lattice study of the
$SU(3)$ model might well lead to phenomenologically significant results.

\acknowledgments{
We thank R. Frezzotti and C. Pena for stimulating comments. The computer
time for the lattice simulations was kindly provided by the University of
Wuppertal.}

\end{document}